\documentclass[%
 reprint,
 amsmath,amssymb,
 aps,
prl,
twocolumns,
]{revtex4-1}
\DeclareSymbolFont{extraup}{U}{zavm}{m}{n}
\DeclareMathSymbol{\vardiamond}{\mathalpha}{extraup}{87}
\usepackage[utf8]{inputenc}
\usepackage{graphicx}
\usepackage{dcolumn}
\usepackage{bm}
\usepackage[french]{babel}

\usepackage{esint}
\usepackage{tikz}

\usepackage{color}

\begin{document}

\preprint{APS/123-QED}

\title{Spin texture driven spintronic enhancement at the LaAlO$_3$/SrTiO$_3$ interface}

\author{Pierre Bruneel}

\author{Marc Gabay}%

\affiliation{%
 Laboratoire de Physique des Solides, CNRS UMR 8502, Université Paris-Saclay, 91405 Orsay Cedex, France
}%

\date{\today}

\begin{abstract}
Recent experiments have shown that transition metal oxide heterostructures such as SrTiO$_3$-based interfaces, exhibit large, gate tunable, spintronic responses. Our theoretical study showcases key factors controlling the magnitude of the conversion, measured by the inverse Edelstein and Spin Hall effects, and their evolution with respect to an electrostatic doping. The origin of the response can be linked to spin-orbital textures.  These stem from the broken inversion symmetry at the interface which produces an unusual form of the interfacial spin-orbit coupling, provided a bulk atomic spin-orbit contribution is present. The amplitudes and variations of these observables are direct consequences of the multi-orbital subband structure of these materials,  featuring avoided and topological crossings. Interband contributions to the coefficients lead to enhanced responses and non-monotonic evolution with doping. We highlight these effects using  analytical approaches and low energy modeling. \end{abstract}

\maketitle


\section*{Introduction}

 Si and GaAs based heterostructures are core elements of our current technology but other approaches are also being pursued. For spintronic devices, one tunes the spin of the carriers to control the charge current \cite{bader_spintronics_2010}. Such conversion is achieved through the Edelstein (spin-galvanic) Effect (EE), where a charge current is converted into a net spin polarization and the Spin Hall Effect (SHE) where a charge current is converted into a spin current. EE has been observed via spin-pumping experiments in semiconductors/ferromagnet heterostructures \cite{dash_electrical_2009,dash_spin_2011,tran_enhancement_2009}, and SHE in bulk transition metals \cite{kimura_room_2007,ryoo_computation_2019, pai_spin_2012, hahn_comparative_2013}, and also in topological insulator heterostructures \cite{bernevig_quantum_2006,konig_quantum_2007}. Both effects and their inverse counterparts occur in systems subject to a Rashba term \cite{sinova_universal_2004,shen_inverse_2014} due to the spin-momentum locking of the electrons. 
Transition Metal Oxide (TMO) heterostructures, notably LaAlO$_3$/SrTiO$_3$ (LAO/STO), have emerged as prime candidates to engineer spintronic devices \cite{ohtomo_high-mobility_2004, gariglio_research_2016, sulpizio_nanoscale_2014}.  
Magnetotransport measurements of STO-based 2DEGs \cite{caviglia_tunable_2010} have helped uncover a large interfacial gate-tunable spin-orbit contribution, referred to as Rashba, ascribed to the confining electric field caused by the discontinuity at the interface. For TMO heterostructures, the multiorbital character of the 2DEG electronic subbands compounded with bulk spin-orbit and inversion symmetry breaking entails a more complex form of spin-orbital-momentum locking \cite{zhong_theory_2013,khalsa_theory_2012}. Large EE efficiencies have been observed in LAO/STO \cite{lesne_highly_2016} or AlOx/STO \cite{vaz_mapping_2019} heterostructures. In addition to displaying a significantly larger signal than in most materials where it is observed, the EE intensity changes with the electrostatic doping. Data analysis of nonlocal transport measurements also suggest that LAO/STO heterostructures host SHE \cite{trier_electric-field_2020}. 

Our study of SrTiO$_3$-based interfaces shows that the  above properties  stem from the particular form of the interfacial spin-orbit coupling which we name  Multiband Interfacial Spin Orbit (MISO). Breaking of inversion symmetry at the interface triggers bond-angle distortions which allow new hoppings between the d-orbitals of the conduction band  and  leads to the MISO term.
Spin-orbital textures emerging from this term strongly impact the magnitude of the EE and SHE responses and their evolution with the doping. 
Analytical results obtained near the bottom of the lowest conduction subband reveal that the amplitudes of the EE and SHE responses are governed by 
the size of the subbands expectation value of the electron spin.
The evolution of the EE term with the filling, when the Fermi energy varies between the bottom of the conduction subband manifold and the first higher-up energy subband (so called Lifshitz transition), is dominated by MISO driven transitions 
 between different subbands (interband spin textures),
except very close to the center of the Brillouin zone (BZ).
Upon further increase of the doping, the anisotropy of the subband dispersions becomes more pronounced and one encounters avoided \cite{king_quasiparticle_2014,zhong_theory_2013,kim_nature_2014} and topological \cite{vivek_topological_2017,sahin_strain_2019} crossings. The size of the EE and SHE reponses depends primarily on changes in nontrivial spin textures and to a lesser extent on the anisotropy of the subband dispersions. The winding orientation of the textures close to the bottom of a subband governs the sign of the Edelstein response for that particular subband. Avoided and topological crossings 
change the ordering of the subbands and hence the nature and the sign of the spin texture windings explaining  the large increase of the interfacial spin-orbit contribution upon doping. Subbands contribute with either positive or negative signs to the EE and SHE terms  causing non uniform variations with respect to electrostatic doping. Except for a very narrow range of wavevectors $k$ near the $\Gamma$ point, 4$\times$4 effective  Hamiltonians are needed to model these features accurately. Numerical computations of these responses using a tight-binding form corroborate the analytical findings and give good agreement with the experiments. 


\begin{figure}
\begin{minipage}[t]{0.53\columnwidth}
\includegraphics[width=1.\linewidth]{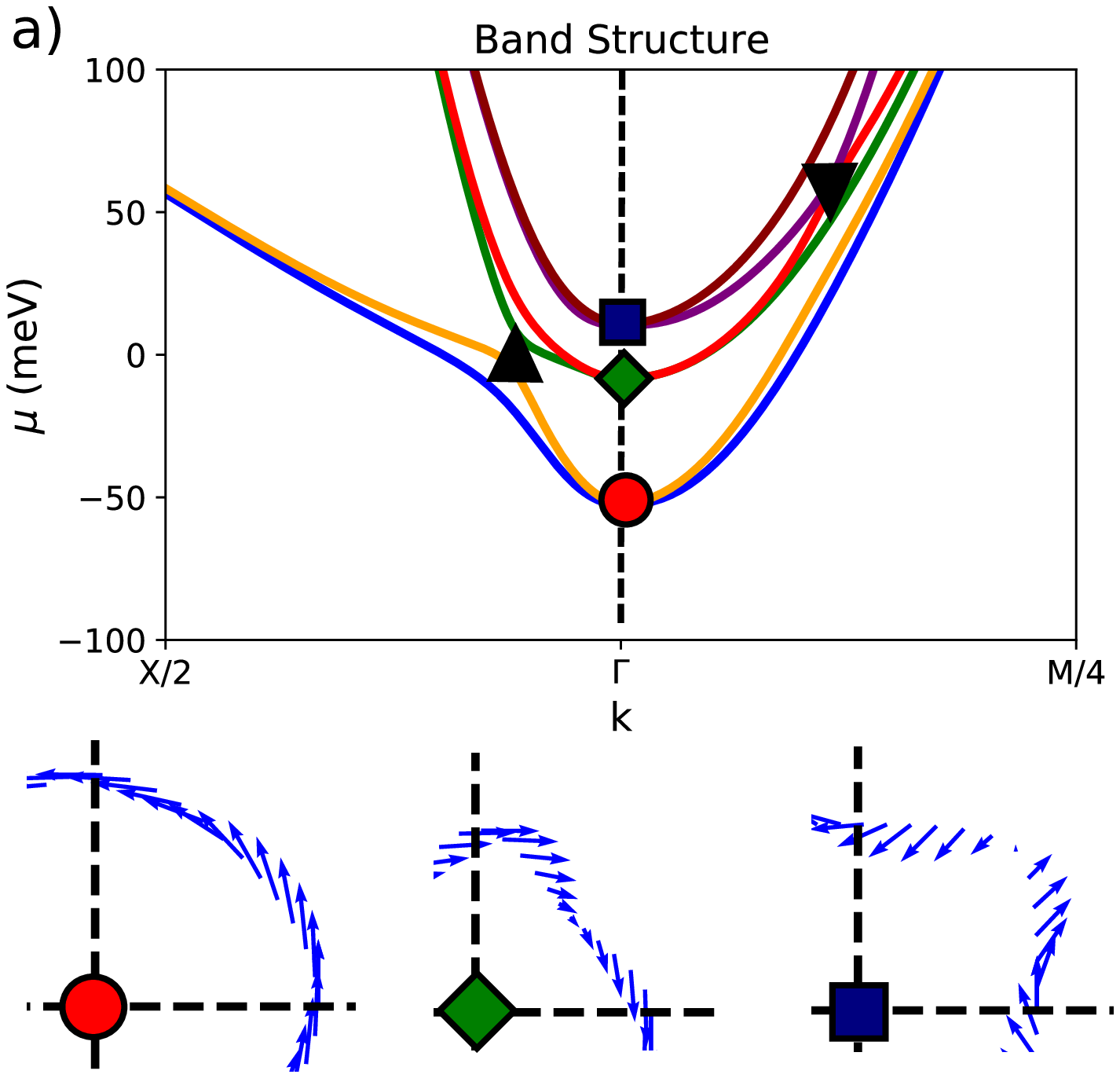}
	\label{fig:sfig1}
\end{minipage}\hfill%
\begin{minipage}[t]{0.47\columnwidth}
\includegraphics[width=1.\linewidth]{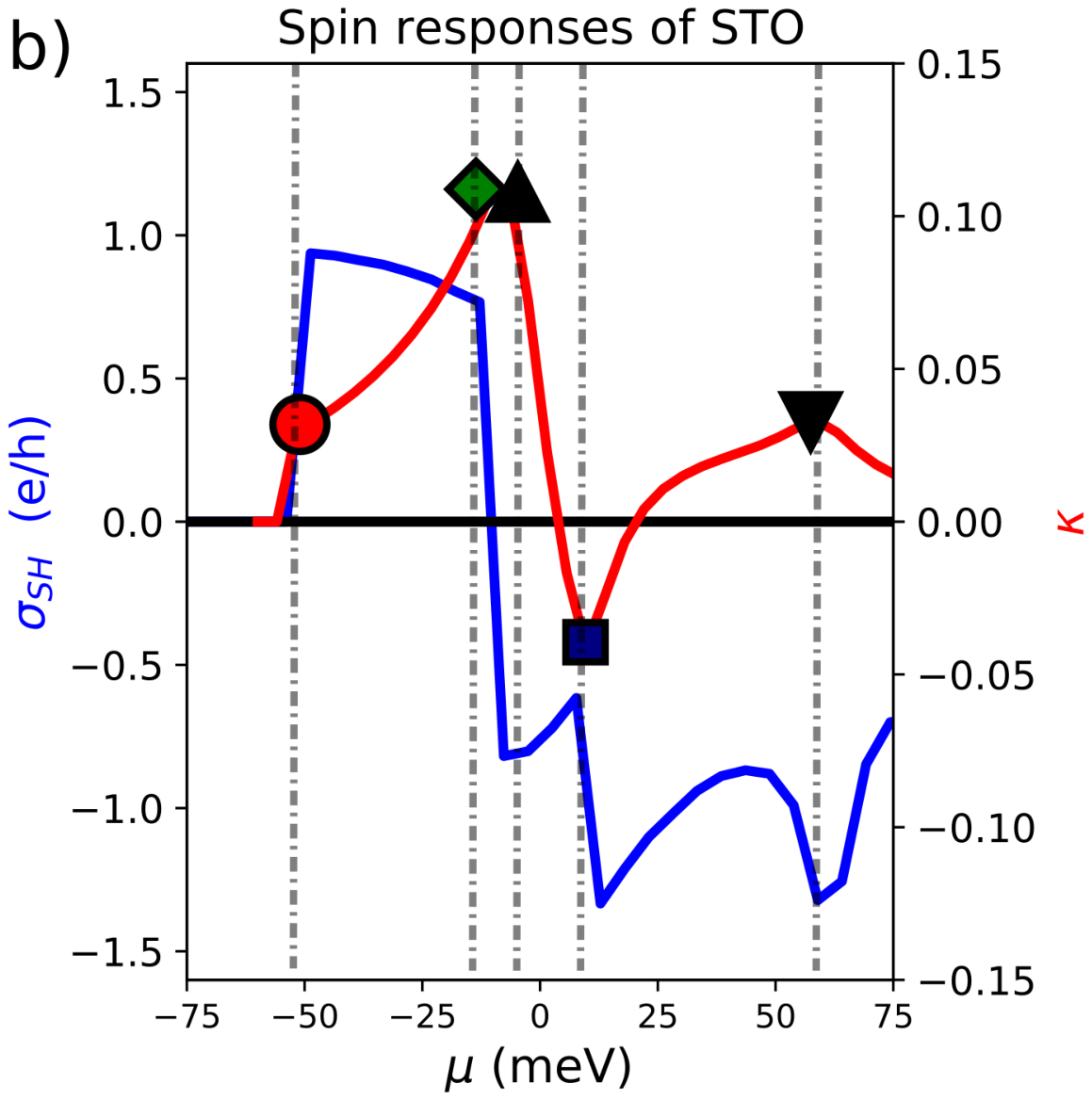}
	\label{fig:sfig2}
\end{minipage}\hfill

\caption{a ) Band structure of the 6-band model  with parameters $t_1$ = 338 meV, $t_2$ = 30 meV, $\Delta$ = 50 meV, $\lambda$ = 8.3 meV, $\gamma_{OM}$ = 20 meV. Path along the X/2-$\Gamma$-M/4 directions in the Brillouin Zone. A (\textcolor{red}{$\Large \bullet$}), B (\textcolor{green}{$\Large \vardiamond$}) and C (\textcolor{blue}{$\Large \blacksquare$}) are located at the bottom of subbands L, N and U. D ($\blacktriangle$) and E($\blacktriangledown$) are avoided and topological crossings. Bottom : spin textures of one Kramers branch near A, B and C for subbands L, N and U (in the first quadrant of the BZ). b ) Variations of the EE (red) and SHE (blue) with the chemical potential. Symbols are the same as for the band structure.}
\label{fig:fig}
\end{figure}
Following \cite{vaz_mapping_2019}, we model the band structure of select STO-based interfaces \cite{harrison_electronic_2012,khalsa_theory_2012,khalsa_theory_2013} using a tight-binding Hamiltonian $\mathcal{H}_{TB}$. It is the sum of three contributions, a kinetic part $\mathcal{H}_{kin}$, a bulk spin-orbit (SOC) part $\mathcal{H}_{SOC}$ and an orbital mixing (OM) part $\mathcal{H}_{OM}$. The detailed form of each term can be found in the tight-binding section of \cite{suppmatSTOSPINTRO} and we will only explain its important features.
Owing to the large crystal field splitting in STO-based interfaces, the relevant conduction band states consist of $t_{2g}$ directional orbitals denoted by $d_{yz},d_{xz},d_{xy}$. For their kinetics, two directions can be labelled as light and one as heavy, with hopping terms $t_1$ and $t_2$ respectively. In the 2DEG, the $d_{xy}$ orbital has a smaller confinement energy than the $d_{yz}$ and $d_{xz}$ orbitals at the $\Gamma$ point. The difference is $\Delta > 0$ \cite{khalsa_theory_2013,harrison_electronic_2012,zhong_theory_2013}.
In \cite{trier_electric-field_2020}, two $d_{xy}$ subbands were used to match the experimental data. Our analytical study only has one, since all the relevant features which determine the characteristics of the EE and SHE responses are found to be adequately captured by the single $d_{xy}$ subband version. We checked that no significant change occurs if we include both $d_{xy}$ subbands for STO or LAO/STO. 
The $t_{2g}$ orbitals are also subject to an atomic (bulk) SOC $\mathcal{H}_{SOC} = \lambda \hat{\vec{L}} \cdot \hat{\vec{\sigma}}$;  
$\hat{\vec{\sigma}}$ are the Pauli matrices  and the orbital operator  $\hat{\vec{L}}$ is expressed in the ($d_{yz},d_{xz},d_{xy}$) basis. Electrostatic imbalance in the vicinity of the surface (or interface) leads to changes in the O-TM-O bond angles such that hoppings between orbitals otherwise forbidden by symmetry are now allowed through orbital mixing \cite{zhong_theory_2013,khalsa_theory_2012,khalsa_theory_2013}. The energy scale of the OM part 
is denoted by $ \gamma_{OM} $.
This term breaks inversion symmetry and is crucial to observe the responses that we discuss below. A compact form, close to the $\Gamma$ point, is $\mathcal{H}_{OM} = 2 \gamma_{OM} a (\hat{\vec{L}} \times \vec{k}) \cdot \vec{z}$ which is a linear Rashba-like form for $\hat{\vec{L}}$. $a$ is the lattice parameter of STO ($a=3.9 \AA$).
The compounded effects of bulk SOC, electrostatic confinement and OM produce the  MISO term.
\vspace {0.25truecm}

Using the symmetries of the system and Stokes theorem, the EE coefficient $\kappa$ is expressed as  (\cite{suppmatSTOSPINTRO})
\begin{equation}
\kappa(\mu) = \frac{e \tau_0}{h} \sum\limits_{\nu} \int\limits^{2\pi}_{0}  k^{\nu}_F(\theta) S^t_{\nu}(\theta) d\theta\\
 = \frac{e \tau_0}{h} \sum\limits_{\nu} \iint\limits_{BZ} d^2k f_{\nu k}(\mu) (\vec{\nabla}_k \times \vec{S}_{\nu} ) \cdot \vec{z} 
	\label{edelresponseFS}
\end{equation}
\begin{figure}
\begin{minipage}[t]{0.49\columnwidth}
  \includegraphics[width=1.\linewidth]{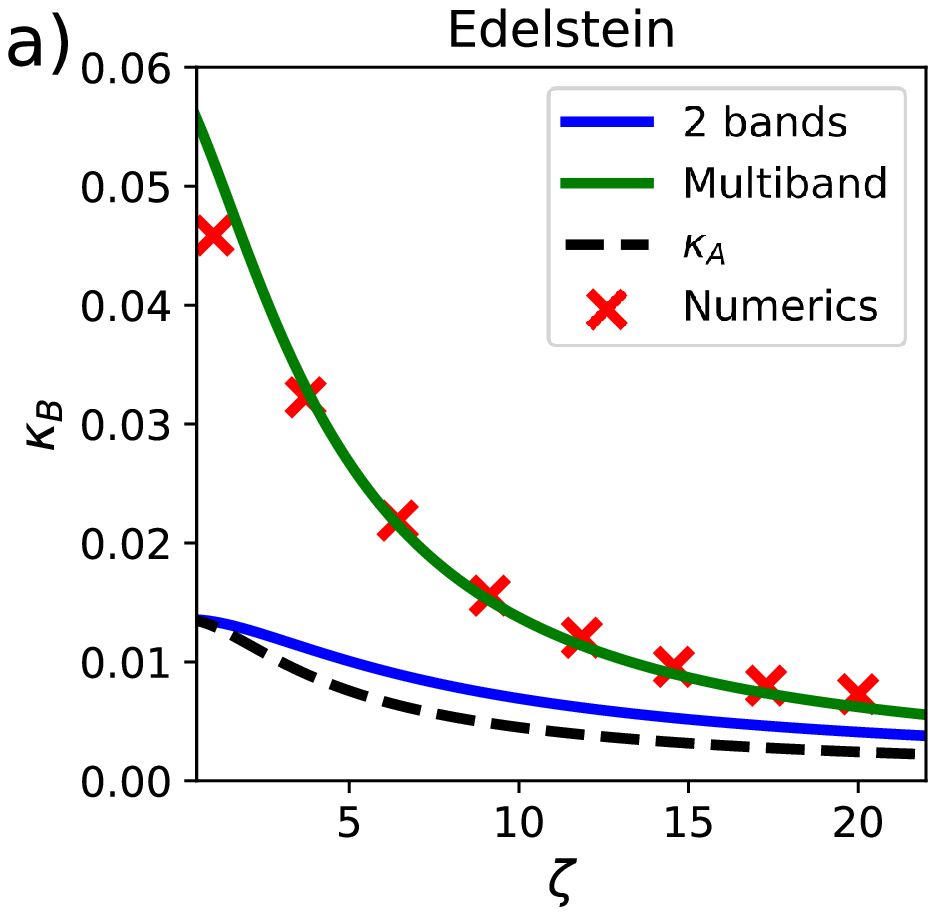}
  \label{fig2:sfig2}
\end{minipage} \hfill
\begin{minipage}[t]{0.49\columnwidth}
  \includegraphics[width=1.\linewidth]{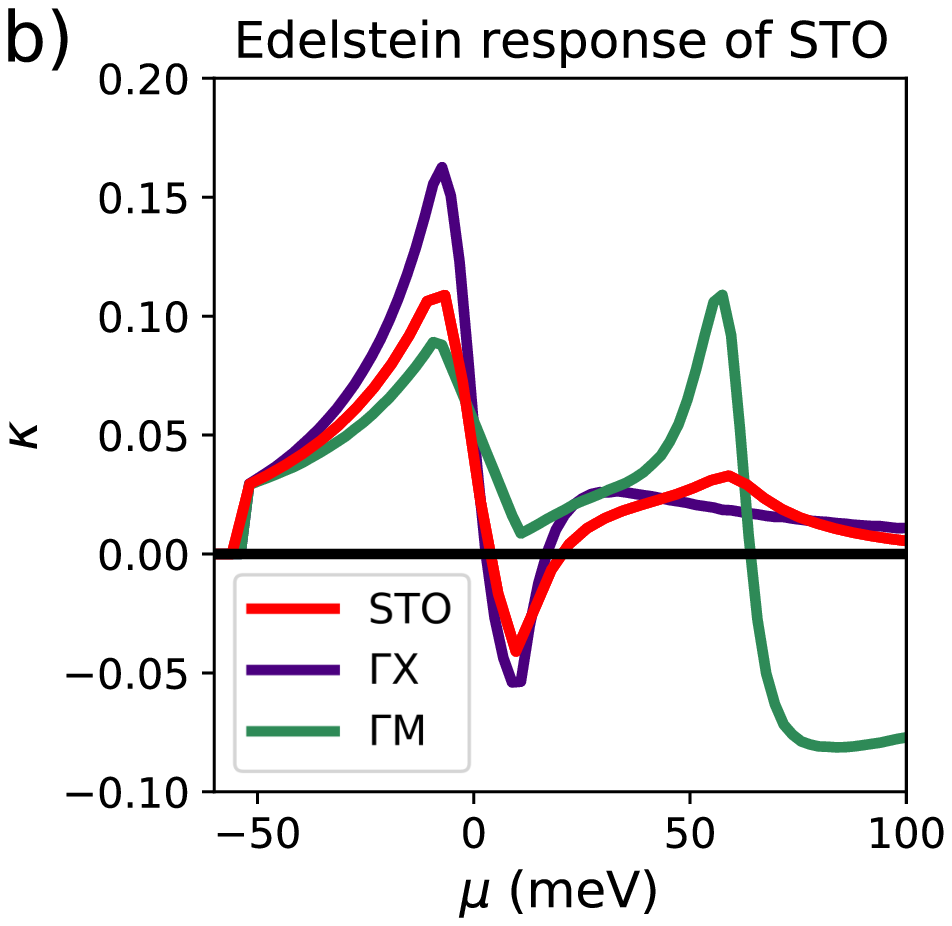}
  \label{fig2:sfig3}
\end{minipage}
\caption{ a ) Variation of the EE $\kappa$ value at A ($\kappa_A$) and at B ($\kappa_B$) with $\zeta = \frac{\Delta}{\lambda}$. Parameters are the same as in Fig. \ref{fig:fig} a, except $\lambda$, which varies. Numerical results (red crosses) are compared with the expression for $\kappa$ in Eq. \ref{Edeltwoparts} : the blue curve is the $\kappa_R$ part of the response and the green curve include all interband corrections. b) Variations of $\kappa$ with $\mu$ from numerical calculations (red),  the $\Gamma$M  (green) and  $\Gamma$X (purple) approximations.}
\label{fig2:fig2}
\end{figure}
$\mu$ is the chemical potential, $k_F^{\nu}(\theta)$ is the Fermi momentum of band $\nu$ at polar angle $\theta$, $S^t_{\nu}$ is the component of the vector field tangential to the Fermi surface of band $\nu$, $f_{\nu k}(\mu)$ is the Fermi function of band $\nu$ and $\tau_0$ is the scattering time (considered to be identical for all the bands). This expression demonstrates the direct connection  between the Edelstein effect, the spin textures and their vorticity.  The relative windings of the textures between the different occupied bands impact the response.
The same formalism yields the Spin Hall Conductivity (SHC), the spin current response to a charge current in the material. For symmetry reasons we only consider the component $\sigma_{SH} = \sigma^z_{xy}$, where z is the direction of polarization of the spin current and $xy$ means that the conductivity is transverse and in the plane of the interface. There are several contributions to this conductivity; a discussion of the extrinsic parts is presented in \cite{suppmatSTOSPINTRO}. Here, we focus on the intrinsic part, which is a coherent property linked to the geometry of the band structure and to interband transitions. The contribution of subband $\nu$ involves the spin curvature $\Omega^{z}_{\nu}(k)$ (see \cite{suppmatSTOSPINTRO}). The total SHC is obtained by summing the contributions over filled states for each band $\nu$.

We diagonalize numerically $\mathcal{H}_{TB}$ . The 
eigenvectors $|\nu,k \rangle$ and corresponding eigenenergies allow us to compute expectation values of any operator A for subband $\nu$ and wavevector k,  $\langle A(k) \rangle_{\nu} = \langle \nu,k | A | \nu,k \rangle$,  and we track their evolution with the chemical potential $\mu$. Note that this procedure assumes a rigid shift of all the bands when applying a voltage bias. This approximation holds for  very low doping, but since the screening properties change with the doping, we should include this effect and recalculate the band structure for each filling; this is done using a Poisson-Schrodinger approach in \cite{suppmatSTOSPINTRO}; we only find differences in prefactors for the responses, but their variations remain unchanged in the experimentally relevant range of fillings.

We now turn to analytical and numerical evaluations of the $\mu$ dependence of the transport coefficients  in the $T=0$ limit. Analytical computations reveal that the EE is proportional to $\gamma_{OM}$ for the experimentally relevant range of values, while the height of the SHC plateaus is not. We emphasize, however, that the presence of the OM term is critical to observe the two responses because it breaks inversion symmetry. 
We also find that the variations of the responses depend sensitively on  $\zeta=\frac{\Delta}{\lambda}$, the balance between confinement and atomic spin-orbit coupling. 
Numerical calculations confirm the above statements. The analytical expressions given below are derived in \cite{suppmatSTOSPINTRO}.
The band structure of the 6-band model for STO is shown in Fig. \ref{fig:fig} a. The orbital gap and atomic SOC lift the degeneracy of the $t_{2g}$ orbitals at the $\Gamma$ point and we label the three subbands  L (lower), N (intermediate) and U (upper). Each subband consists of two Kramers branches. 
We first consider the change in the Edelstein coefficient between the bottom of subbands L and N  (respectively point A and the Lifshitz point B in Fig. \ref{fig:fig} a). As this corresponds to a range where the dominant orbital character is $d_{xy}$, the model is essentially isotropic, implying that $k^{\nu}_F(\theta)$ and $S^t_{\nu}(\theta)$ in Eq. \ref{edelresponseFS} are almost $\theta$-independent and we call them 
$k^{\nu}_{F}$ and $S^{\nu}_{F}$, the effective Fermi wavevector and expectation value of the spin of subband $\nu$, respectively. Hence $\kappa(\mu) \approx  \frac{e \tau_0}{\hbar} \sum\limits_{\nu} k^{\nu}_{F} S^{\nu}_{F}$.
Since subband masses are equal for every orbital along $\Gamma$M, we compute the effective $k^{\nu}_{F}$ and $S^{\nu}_{F}$ using these masses for all $\theta$. We refer to this as the $\Gamma$M approximation and we denote by $\kappa^M$ the value of $\kappa$ that it gives. 
Both the EE and SHE responses display a sharp rise when $\mu$ is at the bottom of  subband L (Fig. \ref{fig:fig} b). Very close to A, these reduce to a linear Rashba-like model since only the Kramers pair within subband L contributes notably and a straightforward calculation yields $\kappa_A \sim \frac{\tau_0 m^{*}}{\hbar^3} \alpha_L S_L$, the product of the Rashba coefficient $\alpha_L= \frac{4}{s} \gamma_{OM} a$, the effective mass $m^{*}$, the scattering time $\tau_0$ and the expectation value of the in-plane spin length $S_L= \frac{1}{2} \left( 1+ \frac{1+\zeta}{s} \right)$ \cite{sinova_universal_2004}. 
$s = \sqrt{\zeta^2+2\zeta+9}$ is the the energy difference between A and C divided by $\lambda$.
Both $\alpha_L$ and $S_L$ vary with $\zeta$ and numerics confirms these analytical forms. 

Analytical computations and low-energy hamiltonian modeling in the vicinity of points A and B of the band structure allow us to probe the variations of the EE and SHE responses in the vicinity of other special points of the band structure 
and from the study of the EE response between A and B we are able to explain the behavior of the Edelstein response for the relevant values of $\mu$ ; experiments and numerics show that the amplitude of the EE term rises sharply in this energy range. For a linear Rashba Hamiltonian such that both Kramers branches are populated we expect a $\mu$-independent value of the EE coefficient but we only see this very close to A. The value of $\kappa$  for STO more than doubles between A and B and the amplitude of the peak at B decreases with $\zeta$ (Fig. \ref{fig2:fig2}). 
If the only contributions to the EE came from the L subband, we would expect the variation of $\kappa$ to be caused by $\zeta$, which depends on the confinement gap $\Delta$ ; as $\mu$ increases, the energy difference between subbands L and U (the effective $\Delta(\mu)$) decreases and the anisotropy of the dispersions in k space become pronounced as point B is approached.
While this gives corrections to $\alpha_L$ and $S_L$, it fails to account for the value $\kappa_B$ of $\kappa$ at point B (Fig. \ref{fig2:fig2}). One needs to include corrections to the expectation value of the spin originating from transitions to the higher energy subbands (N, U) which occur because of the MISO form of the interfacial spin-orbit term. The expectation value of $S_L$ for the Kramers pair $\pm$ of subband L  becomes $S^{\pm}_L = \mp S_0(\zeta,\theta) + \delta S(\zeta,\theta)$ where $\theta$ is the angle of the k vector, and from Eq. \ref{edelresponseFS} the EE response for the Kramers partners takes the form :
\begin{equation}
\kappa = \kappa_R+\kappa_{MB} = \frac{e \tau_0}{h} \int\limits_0^{2\pi} \left[ \delta k^F_{\theta} S_0(\zeta,\theta) + k^F_{\theta} \delta S(\zeta,\theta) \right] d\theta
\label{Edeltwoparts}
\end{equation}
$\delta k^F_{\theta}$ is the splitting in k for the Kramers pair $\pm$ at $k^F_{\theta}$. $\kappa_R$ stands for Rashba-like and $\kappa_{MB}$ for Multiband. For the L subband as stated before we can ignore the $\theta$ dependence  as the system is essentially isotropic and we get  $\kappa \approx \kappa^M$. $\kappa_R$ is the main contribution very close to A but it is dominated by $\kappa_{MB}$  upon increasing the filling even slightly, because then $\delta k << k$. $\kappa_{MB}$ is a distinctive signature of the MISO term. The Rashba-like winding textures of  $S_0$ that one gets for subband L  can be called \textit{intraband}, and the textures of $\delta S$ resulting from matrix elements between subbands L and N , L and U and N and U, \textit{interband}. Contributions due to the coupling of L to N and U are encoded in $\kappa_{MB}$. Within each subband, spin textures of the time reversed Kramers branches wind in opposite direction, as expected. However the winding in each branch is a superposition of two patterns : the intraband described by $S_0$, and the interband described by $\delta S$. For the two Kramers partners, the intraband pattern gives a winding in opposite directions  while the interband pattern gives a winding in the same direction. We stress that this decomposition of the EE response in two parts is valid for all the subbands.
For $\mu$ between A and B, Eq. \ref{Edeltwoparts} can be evaluated analytically using the $\Gamma$M approximation and cast in the form
\begin{equation}
\kappa = \frac{e \tau m^*}{\hbar^3} \gamma_{OM} a g(\zeta,\mu)
	\label{Edelscaling}
\end{equation}
$\tau$ is the scattering time, $m^*$ the effective mass and $g(\zeta,\mu)$ is a band structure dependent factor taking into account the intraband and interband effects. 
This gives an accurate estimate of the evolution of the EE response for a wide range of dopings between A and B. Figs. \ref{fig2:fig2} a and $\;\;\;$ 2 b compare the numerically determined  variations of $\kappa_B$  with $\zeta$ and of $\kappa$ with $\mu$ along with the analytical expression of $\kappa^M$, for the parameters of STO. Numerical and experimental data agree well.
Eq. \ref{Edelscaling} may be interpreted as the Rashba-Edelstein response of a system with an effective Rashba coefficient $\alpha_{eff} = \gamma_{OM}a g(\zeta,\mu)$ that depends on $\mu$. It is commonly stated that the LAO/STO interface is subject to an effective Rashba coupling the intensity of which varies strongly with the chemical potential in the vicinity of the avoided band gaps of the band structure \cite{zhong_theory_2013}. The approach developped throughout this paper points to an alternative explanation where spin textures play a prominent role. The spin-orbital structure of the eigenfunctions is critical to explain the variations of $\alpha_{eff}$ with $\mu$.


Eqs. \ref{edelresponseFS} shows that $\kappa$ is a sum of contributions coming from the different subbands. We now turn to the behavior of the system at higher fillings (Fig. \ref{fig:fig} b).  Based on the above analysis, we discuss the evolution of the EE response  in terms of spin textures, as $\mu$ sweeps through other remarkable points of the band structure (B, C, D and E).
We start with a perturbative analysis, between B and C,  close to point B ; for these fillings, corrections due to the anisotropic dispersion can no longer be neglected. We project out the L subband  (\cite{suppmatSTOSPINTRO}),  and get a 4$\times$4 effective hamiltonian giving the electronic spectrum of the N and U levels. Close to $\Gamma$, it further reduces to an effective 2-band model for subband N. The energy splitting between the doublet in subband N is of order $k^3$, but one should not conclude that we get a simple cubic Rashba term. The expectation values of the spins, which wind around the Fermi contours yielding the intraband spin textures of N and U, change around the contour. Eq. \ref{Edeltwoparts} can be evaluated and contrary to what we found for subband L near point A, the EE response here is dominated by the interband contributions as the intraband contribution vanishes to first order. The amplitude of this contribution is larger for $\theta$ close to 0 (the X direction); Fig. \ref{fig2:fig2} b shows that this $\Gamma$X approximation for $\theta$ gives a close match to the numerical determination of the EE near B. Note that the spin textures of the N subband wind in a direction which is opposite to that of the L subband : this causes a downturn in the variation of $\kappa$ with $\mu$ beyond B.
As $\mu$ exceeds the energy corresponding to C, subband U become filled. The dispersion is anisotropic (square-like Fermi contour)and the spin textures correspond to an anti-vortex with $\pi$-domains in the $\Gamma$M direction, and the winding is in the same direction as for subband L, as depicted in Fig. \ref{fig:sfig1}1. This explains the upturn of the curve beyond C (Fig. \ref{fig:fig} b).
In the vicinity of the avoided (D) and topological (E) crossings, the ordering of the bands changes. Couplings affect the spin textures strongly, leading to plateaus for $\kappa$ near these points. It is however less straightforward to write effective models similar to those we could construct near B and C.


The evolution of the SHC with doping can be discussed in similar terms. With increasing filling, new subbands and spin textures contribute to the spin-to-charge conversion as $\mu$ passes through the special points that we identified above. The intensities of the SHE term for the L and U subbands is correctly captured using the same modeling and derivation that we used to compute the EE response (see \cite{suppmatSTOSPINTRO}). Intraband transitions give  the response of a Linear Rashba-like model albeit with only a partial projection of the spin; in units of $\frac{e}{h}$, the height of the intrinsic SHC plateau at A is equal to the expectation value of $S_z$ for the Kramers pair in subband L,  $S_z^L = \frac{\zeta+1}{s}$. For a pure Rashba energy dispersion, we would get a quantized value of one. Our numerical computations confirm this.
When SOC is small compared to the confinement gap, the value of the z component of the spin is largest, but when the two are of the same order of magnitude, the mixed orbital character of the states causes a reduction of the spin projection and of the efficiency of the conversion. This intrinsic conversion can also be affected by disorder (\cite{suppmatSTOSPINTRO}); in the large $\zeta$ limit, the Rashba energy is lower and disorder should impact the amplitude of the conversion more severely. Proper account of vertex corrections would quantify this,  but this is beyond the scope of our study. The contribution of subband U is opposite to that of  subband L because the sign of the z component of the expectation value of the spin is opposite for the two subbands.

For subband N, the intraband part is zero if the anisotropic dispersion is not taken into account. Perturbation theory predicts a value  $-1$ for the height of the SHC plateau. The sign is correct  and so is the amplitude close to B. The pattern of the energy dispersions becomes complex as $\mu$ varies from the Lifshitz transition to the avoided crossings. Other interband contributions emerge and the evolution of the plateaus depends on the parameters of the tight binding Hamiltonians of STO and LAO/STO.

Numerical analysis shows that the contributions of the various subbands are additive provided subbands of a given orbital character do not couple;  including a second $d_{xy}$ subbband to the Hamiltonian, as was done in Refs \cite{vaz_mapping_2019, trier_electric-field_2020} amounts to simply shift up the EE and SHE curves. Note that disorder affects the lowest energy subband most so that its contribution might not even be seen. Disorder also impacts the amplitude of the responses but far less their variations with the filling. At very low dopings, the response simply depends  on  the effective mass, the amplitude of the spin-orbit coupling and the strength of the confinement.\\
We established that a model of $t_{2g}$ orbitals can account for the Edelstein and Spin Hall responses which have been observed in TMO heterostructures. Key to these spintronic observables are the presence of atomic SOC and orbital mixing causing spin-orbital momentum locking and winding spin textures in the electronic subbands.  Contributions from the various subbands add up, but they do not all wind in the same direction causing non uniform variations of the EE and SHE coefficients upon doping.  The multiorbital character of the states  becomes more pronounced as the filling increases and as one moves away from the $\Gamma$ point. Spin texture contributions due to transitions between the Kramers branches of the same subband (intraband) are Rashba-like. They are interband when pertaining to transitions between subbands split by an energy amount larger than the orbital mixing. Our study underscores the significance of  multiband physics in explaining the strong deviations of the spin-to-charge conversion coefficients from their standard Rashba behavior. We can use it to make predictions about other types of perovskites; the electrostatic confinement needs to be of the same order of magnitude as the atomic spin-orbit coupling in order to maximize the spin-to-charge conversion. For instance, we estimate that for KTaO$_3$ the geometric factor (Eq. \ref{Edelscaling}) is  2-3 times larger than for STO suggesting an enhanced value of $\kappa$.

We would like to thank Manuel Bibes and Agn\`es Barth\'el\'emy for insightful discussions. MG gratefully acknowledges support from the French National Research Agency (ANR) as part of the  ``Investissements d'Avenir program'' (Labex NanoSaclay, ANR-10-LABX-0035).

\bibliographystyle{unsrt}
\bibliography{mybibmain}

\begin{thebibliography}{10}

\bibitem{bader_spintronics_2010}
S.D. Bader and S.S.P. Parkin.
\newblock {\em Annual Review of Condensed Matter Physics}, 1(1):71--88, August
  2010.

\bibitem{dash_electrical_2009}
Saroj~P. Dash, Sandeep Sharma, Ram~S. Patel, Michel~P. de~Jong, and Ron Jansen.
\newblock {\em Nature}, 462(7272):491--494, November 2009.

\bibitem{dash_spin_2011}
S.~P. Dash, S.~Sharma, J.~C.~Le Breton, H.~Jaffrès, J.~Peiro, J.-M. George,
  A.~Lemaître, and R.~Jansen.
\newblock {\em Physical Review B}, 84(5):054410, August 2011.

\bibitem{tran_enhancement_2009}
M.~Tran, H.~Jaffrès, C.~Deranlot, J.-M. George, A.~Fert, A.~Miard, and
  A.~Lemaître.
\newblock {\em Physical Review Letters}, 102(3):036601, January 2009.

\bibitem{kimura_room_2007}
T.~Kimura, Y.~Otani, T.~Sato, S.~Takahashi, and S.~Maekawa.
\newblock {\em Physical Review Letters}, 98(15):156601, April 2007.

\bibitem{ryoo_computation_2019}
Ji~Hoon Ryoo, Cheol-Hwan Park, and Ivo Souza.
\newblock {\em Physical Review B}, 99(23):235113, June 2019.

\bibitem{pai_spin_2012}
Chi-Feng Pai, Luqiao Liu, Y.~Li, H.~W. Tseng, D.~C. Ralph, and R.~A. Buhrman.
\newblock {\em Applied Physics Letters}, 101(12):122404, September 2012.
\newblock Publisher: American Institute of Physics.

\bibitem{hahn_comparative_2013}
C.~Hahn, G.~de~Loubens, O.~Klein, M.~Viret, V.~V. Naletov, and J.~Ben~Youssef.
\newblock {\em Physical Review B}, 87(17):174417, May 2013.
\newblock Publisher: American Physical Society.

\bibitem{bernevig_quantum_2006}
B.~Andrei Bernevig, Taylor~L. Hughes, and Shou-Cheng Zhang.
\newblock {\em Science}, 314(5806):1757--1761, December 2006.

\bibitem{konig_quantum_2007}
Markus König, Steffen Wiedmann, Christoph Brüne, Andreas Roth, Hartmut
  Buhmann, Laurens~W. Molenkamp, Xiao-Liang Qi, and Shou-Cheng Zhang.
\newblock {\em Science}, 318(5851):766--770, November 2007.

\bibitem{sinova_universal_2004}
Jairo Sinova, Dimitrie Culcer, Q.~Niu, N.~A. Sinitsyn, T.~Jungwirth, and A.~H.
  MacDonald.
\newblock {\em Physical Review Letters}, 92(12), March 2004.

\bibitem{shen_inverse_2014}
Ka~Shen, G.~Vignale, and R.~Raimondi.
\newblock {\em Physical Review Letters}, 112(9), March 2014.

\bibitem{ohtomo_high-mobility_2004}
A.~Ohtomo and H.~Y. Hwang.
\newblock {\em Nature}, 427(6973):423--426, January 2004.

\bibitem{gariglio_research_2016}
S.~Gariglio, M.~Gabay, and J.-M. Triscone.
\newblock {\em APL Materials}, 4(6):060701, June 2016.

\bibitem{sulpizio_nanoscale_2014}
Joseph~A. Sulpizio, Shahal Ilani, Patrick Irvin, and Jeremy Levy.
\newblock {\em Annual Review of Materials Research}, 44(1):117--149, 2014.

\bibitem{caviglia_tunable_2010}
A.~D. Caviglia, M.~Gabay, S.~Gariglio, N.~Reyren, C.~Cancellieri, and J.-M.
  Triscone.
\newblock {\em Physical Review Letters}, 104(12):126803, March 2010.

\bibitem{zhong_theory_2013}
Zhicheng Zhong, Anna Tóth, and Karsten Held.
\newblock {\em Physical Review B}, 87(16):161102, April 2013.
\newblock Publisher: American Physical Society.

\bibitem{khalsa_theory_2012}
Guru Khalsa and A.~H. MacDonald.
\newblock {\em Physical Review B}, 86(12):125121, September 2012.

\bibitem{lesne_highly_2016}
E.~Lesne, Yu~Fu, S.~Oyarzun, J.~C. Rojas-Sánchez, D.~C. Vaz, H.~Naganuma,
  G.~Sicoli, J.-P. Attané, M.~Jamet, E.~Jacquet, J.-M. George,
  A.~Barthélémy, H.~Jaffrès, A.~Fert, M.~Bibes, and L.~Vila.
\newblock {\em Nature Materials}, 15(12):1261--1266, December 2016.

\bibitem{vaz_mapping_2019}
Diogo~C. Vaz, Paul Noël, Annika Johansson, Börge Göbel, Flavio~Y. Bruno,
  Gyanendra Singh, Siobhan McKeown-Walker, Felix Trier, Luis~M. Vicente-Arche,
  Anke Sander, Sergio Valencia, Pierre Bruneel, Manali Vivek, Marc Gabay,
  Nicolas Bergeal, Felix Baumberger, Hanako Okuno, Agnès Barthélémy, Albert
  Fert, Laurent Vila, Ingrid Mertig, Jean-Philippe Attané, and Manuel Bibes.
\newblock {\em Nature Materials}, pages 1--7, September 2019.

\bibitem{trier_electric-field_2020}
Felix Trier, Diogo~C. Vaz, Pierre Bruneel, Paul Noël, Albert Fert, Laurent
  Vila, Jean-Philippe Attané, Agnès Barthélémy, Marc Gabay, Henri Jaffrès,
  and Manuel Bibes.
\newblock {\em Nano Letters}, 20(1):395--401, January 2020.
\newblock Publisher: American Chemical Society.

\bibitem{king_quasiparticle_2014}
P.~D.~C. King, S.~McKeown Walker, A.~Tamai, A.~de~la Torre, T.~Eknapakul,
  P.~Buaphet, S.-K. Mo, W.~Meevasana, M.~S. Bahramy, and F.~Baumberger.
\newblock {\em Nature Communications}, 5(1):3414, May 2014.
\newblock arXiv: 1403.0520.

\bibitem{kim_nature_2014}
Panjin Kim, Kyeong~Tae Kang, and Jung~Hoon Han.
\newblock {\em arXiv:1404.6858 [cond-mat]}, April 2014.

\bibitem{vivek_topological_2017}
Manali Vivek, Mark~O. Goerbig, and Marc Gabay.
\newblock {\em Physical Review B}, 95(16):165117, April 2017.
\newblock Publisher: American Physical Society.

\bibitem{sahin_strain_2019}
Cüneyt Şahin, Giovanni Vignale, and Michael~E. Flatté.
\newblock {\em Physical Review Materials}, 3(1):014401, January 2019.
\newblock Publisher: American Physical Society.

\bibitem{harrison_electronic_2012}
Walter~A. Harrison.
\newblock Courier Corporation, March 2012.

\bibitem{khalsa_theory_2013}
Guru Khalsa, Byounghak Lee, and A.~H. MacDonald.
\newblock {\em Physical Review B}, 88(4):041302, July 2013.

\bibitem{suppmatSTOSPINTRO}
See the supplementary material for details about the tight binding model the
  details of the derivation of~the intraband and interband formulas.

\end{thebibliography}

\end{document}